\documentclass{aa}  
\usepackage{amssymb}
\usepackage{graphicx}
\usepackage[authoryear]{natbib}
\usepackage[figuresright]{rotating}
\usepackage{color}

\usepackage{txfonts}
\newcommand{\teff}{\ifmmode T_{\rm eff} \else T$_{\mathrm{eff}}$~\fi}
\newcommand{\logg}{\ifmmode \log g \else $\log g$~\fi}
\newcommand{\lL}{\ifmmode \log(L/L_{\odot}) \else $\log(L/L_{\odot})$~\fi}

\newcommand{\vsini}{$V$ sin$i$}

\newcommand{\kms}{km s$^{-1}$~}
\newcommand{\msun}{\ifmmode M_{\odot} \else M$_{\odot}$~\fi}
\newcommand{\zsun}{\ifmmode Z_{\odot} \else Z$_{\odot}$~\fi}
\newcommand{\lsun}{\ifmmode L_{\odot} \else L$_{\odot}$~\fi}
\newcommand{\rsun}{\ifmmode R_{\odot} \else R$_{\odot}$~\fi}
\newcommand{\qh}{\ifmmode Q_{\rm H} \else $Q_{\rm H}$~\fi}
\newcommand{\ciso}{\ifmmode ^{12}{\rm C}/^{13}{\rm C} \else $^{12}{\rm C}/^{13}{\rm C}$~\fi}
\newcommand{\qhei}{\ifmmode Q_{\ion{He}{i}} \else $Q_{\ion{He}{i}}$\fi}

\newcommand{\mum}{\ifmmode \mu m \else $\mu m$\fi}

\begin{document}

\title{Weak G-band stars on the H-R Diagram: \\ Clues to the origin of
Li anomaly}

   \author{
          A. Palacios\inst{1} 
          \and
          M. Parthasarathy\inst{2,3,4}
          \and
          Y.Bharat Kumar\inst{5}
          \and
          G. Jasniewicz\inst{1}
          }

   \offprints{A. Palacios: ana.palacios AT univ-montp2.fr}

   \institute{LUPM UMR 5299 CNRS/UM2, Universit\'e Montpellier II, CC 72, 34095 Montpellier Cedex 05, France
         \and
             National Astronomical Observatory of Japan 2-21-1 Osawa, Mitaka, Tokyo 181-8588, Japan
         \and
             McDonnell Center for the Space Sciences, Physics department,
         Washington University in St. Louis, One Brookings Drive St. Louis, MO 63130, USA
         \and
             Aryabhatta Research Institute of Observational Sciences
         (ARIES) Nainital - 263129, India
         \and
             Indian Institute of Astrophysics, 2nd block Koramangala, Bangalore, 560034 India
             }

   \date{}

\authorrunning{A. Palacios et al.}
\titlerunning{Weak G-band stars on the H-R Diagram}

\date{Received, accepted}

\abstract
   {}
  {Weak G-band (WGB) stars are a rare class of cool luminous stars that
  present a strong depletion in carbon, but also lithium abundance anomalies that have been little explored in
  the literature since the first discovery of these peculiar objects in
  the early 50's. Here we focus on the Li-rich WGB stars and report on
  their evolutionary status. We explore different paths to propose a
  tentative explanation for the lithium anomaly. }
  {Using archive data, we derive the fundamental parameters of WGB
  (\teff, \logg, \lL) using Hipparcos parallaxes and recent temperature
  scales. From the equivalent widths of Li resonance line at 6707 \AA,
  we uniformly derive the lithium abundances and apply when possible
  NLTE corrections following the procedure described by Lind et al. (2009). We also compute dedicated stellar evolution models in the
  mass range 3.0 to 4.5 \msun, exploring the effects of rotation-induced
  and thermohaline mixing. These models are used to locate the WGB stars
  in the H-R diagram and to explore the origin of the abundance
  anomalies.}
  { The location of WGB stars in the H-R diagram shows that these are
  intermediate mass stars of masses ranging from
3.0 to 4.5 M$_{\odot}$ located at the clump, which implies
  a degeneracy of their evolutionary status between subgiant/red giant
  branch and core helium burning phases. The atmospheres of
a large proportion of WGB stars (more than 50\%) exhibit lithium
  abundances $A(Li) \geq 1.4$ dex similar to Li-rich K giants. 
The position of WGB stars along with the Li-rich K giants in the H-R
diagram however indicates that both are well separated groups. The
  combined and tentatively consistent analysis of the abundance pattern
  for lithium, carbon and nitrogen of WGB stars seems to indicate that carbon
  underabundance could be decorrelated from the lithium and nitrogen
  overabundances.}
  {}


\maketitle

\section{Introduction}\label{sec:intro}
\subsection*{ Defining WGB stars}
 The weak G-band (hereafter WGB) stars are G and K giants whose spectra show very weak or
absent G-bands of the CH A$^{2}\Delta$ - X$^{2}\Pi$ system at 4300\AA. These
stars were first identified as stellar class by \citet{bidelman1951},
and have been mainly studied in the late seventies, early eighties with
a total number of dedicated papers not exceeding 20 \citep[see e.g. for
instance][]{sneden1978,rao1978,partha1980,day1980}. They are rare with
less than 30 known to date among the population of G-K giants in the Galaxy.\\

Chemical composition studies \citep{sneden1978,cottrell1978}
 demonstrated that they are very much underabundant in carbon (typical
 [C/Fe] $\approx -1.4$) and
 present small overabundances of nitrogen and normal oxygen. \citet{sneden1978}
 analyzed the $^{13}$CN red-system features in the high resolution spectra
of weak G-band stars and found \ciso = 4.  \citet{hartoog1977} from the
photometry of the 2.3$\mu$m CO vibration-rotation bands confirmed the
underabundances of carbon in a large sample of weak G-band stars.
The CH band strengths observed by \citet{rao1978} also indicate that the
weakening of the G-band is due to underabundance of carbon.
\citet{sneden1984} obtained high resolution spectra of the
2$\mu$m first-overtone CO bands in the weak G-band giants and found
excellent agreement between the carbon abundances derived from CO data
and those determined using features of the CH G-band. They have found
\ciso $\le$ 4  which is in agreement with the 
predicted ratio for the CN-cycle in equilibrium. 
The normal abundance of oxygen and sodium shows that the atmospheres of
WGB stars are probably not mixed with ON-cycle processed material \citep{sneden1978,drake1994}.\\ 

Additional chemical constraints are given by lithium and and beryllium abundance determinations.\\ Li abundances have been derived for several WGB
stars and several stars were found to be lithium rich, e.g.  with
A(Li)\footnote{With the classical notation A(Li) =
log(${\rm N}_{Li}/{\rm N}_H) + 12$}
$\ge 1.4$ \citep[see][]{brown1989} and reaching up to  A(Li) = 3
\citep{lambert1984,partha1980}. \citet{lambert1984} also report on the
possible presence of $^6{\rm Li}$ in HR 1299, but the profile of the
6707 \AA $^7{\rm Li}$ line that is better reproduced with $^6{\rm
Li}/^7{\rm Li} \ne 0$, could also be reproduced without having to invoke
 the presence of $^6{\rm Li}$. \\Be abundances were derived from IUE
 spectra for 3 Li-rich weak G-band stars by \citet{partha1984}. Providing a
 differential analysis between the WGB stars and K giants in the Hyades,
 Be is found to be similar in both groups, and compatible within
 presumably large errorbars\footnote{The authors do not indicate any
 errorbar in the paper and give abundances of Be ``{\em rounded to the
 nearest  0.5 dex}''.}, with the expected post dredge-up Be abundances
 according to standard stellar evolution models.

\subsection*{Tentative evolutionary status and initial mass estimate}

 Due to their relative low temperature and gravity, WGB stars are
classified as giants and fall amidst the very crowded area of the HR
diagram populated with stars on the subgiant branch, the RGB, the core
helium burning phase and the AGB. This has made the definition of their
evolutionary status and initial mass difficult. They seem however to
have been unanimously qualified as core helium burning star. In
particular \citet{sneden1978} tentatively attribute a mass of $\approx 1
\msun$ to HR 6766 (e.g. HD 165634) and exclude it from being more
massive than 3 \msun because of the magnitude they derive. With such a
low-mass, that star should then have undergone the He flash. Let us note
that their conjecture is based on a very crude estimation for the
luminosity due to the large error bars impairing the parallax of HR
6766. \citet{cottrell1978} also identify WGB stars in their sample as
being in the mass range 1.5 \msun $\leq$ M $\leq$ 3 \msun past the He
flash. Later on, \citet{lambert1984} proposed that WGB stars could be
the progeny of magnetic Ap stars, with initial masses of 2-3 \msun that
have undergone a dredge-up, but they do not make any clear statement of
their evolutionary status. Much recently \citet{partha2000,partha2002} suggest that
WGB stars should be the descendants of intermediate-mass stars with 2
\msun $\leq$ M $\leq$ 5 \msun. In that case, WGB should not go through
the He flash, and will undergo a different evolution than that of the
lower mass stars. This in turn may have an impact on the possible
mechanisms that can be invoked to explain the abundance anomalies. 

\subsection*{Comparison with other carbon depleted and Li-rich giants}

 WGB stars are thus possibly low to intermediate mass evolved stars exhibiting strong depletion of carbon that can
be accompanied in several cases by a Li enrichment.
One may wonder how these objects compare with the other stars that also
present signatures of CN-cycle processed material in their atmospheres
and exhibit unexpected Li enrichment.\\
Giant stars with depleted carbon and enhanced nitrogen abundances are
found in almost all environments among giants. These are essentially low-mass stars
on the RGB. \citet{charbon1998}
show that 96\% of field RGB stars with metallicities from solar to
1/1000 solar and M$_V \leq -0.5$ have a carbon isotopic ratio lower than
what is predicted by standard theory. 
In their survey of metal-poor field stars, \citet{gratton2000} show that
the unexpected decrease of the carbon isotopic ratio occurs at the RGB
bump, when the outgoing hydrogen burning shell (hereafter HBS) erases
the mean molecular weight discontinuity left behind by the retreating
convective envelope as a result of the first dredge-up. It is
furthermore associated with a decrease of the carbon abundance [C/Fe], an increase of the
nitrogen abundance [N/Fe] and a strong decrease of the lithium
abundance, all characteristic of CN-cycle processed material. Oxygen
does not seem to be affected, as in the case of WGB stars\\%
Typically, \citet{charbon1998} report that around solar
metallicity the lower envelope for the \ciso ratio in low-mass RGB stars
is of about 12. \citet{gratton2000} report  [C/Fe] = -0.58 $\pm$ 0.03 as a mean
carbon abundance for upper RGB low-metallicity field stars (e.g. \lL $>$ 2
and -2 $\le$ [Fe/H] $\leq$ -1), and \ciso between 6-10. These values
are higher than the mean obtained for WGB stars.\\
In addition to that Li-enrichment also exists among K giants. It is a
very rare phenomenon found in less than 2\% of the K giant stars
\citep{kumar2011}. \citet{charbon2000} have shown that around solar
metallicity, giants exhibiting
a genuine overabundance of Li (e.g. excluding those giants that are
still undergoing the first dredge-up) appear to cluster around the RGB
bump and the clump. Very recently however \citet{monaco2011} have studied
Li-rich RGB stars in the Galactic thick disk, and show that at low
metallicity Li-rich giants do not cluster at the bump and are found
scattered between the bump and the tip of the RGB \citep[see also][]{lebzelter2011}. \citet{kumar2011} confirmed
the finding by \citet{charbon2000}, with Li-rich stars clustered at the
RGB bump, and some new objects that seem to be more evolved, may be past
the RGB tip and undergoing central He burning. Let
us note that Li-rich K giant stars around the bump region do not present ``anomalous'' carbon
isotopic ratios, e.g., the \ciso determined in these stars well agrees
with the predictions of post dredge-up
dilution \citep{charbon2000,kumar2011}.\\
In these K giants, Li enrichment is thought to have an internal origin
\citep[so called][mechanism]{cf1971} and to be a short-lived phase {\em
preceding} the phase of transport CN-cycle processed material from the
HBS to the surface \citep{charbon2000,palacios2001}. The extra-mixing mechanism capable of transporting Li in
regions where it is preserved from proton captures and connecting the
convective envelope to these regions is not clearly identified \citep{palacios2001,palmerini2011}.

As a summary, in low-mass RGB stars, extra-mixing processes are clearly responsible for
bringing up to the stellar surface the products of the CN-cycle occurring
in the outer layers of the HBS deeper inside the star, and may be also
responsible for strong and short-lived episodes of Li enrichment. The
specific location of Li-rich K giants \citep[see, for
example][]{jasniewicz1999} points towards an extra-mixing
process that becomes efficient only when no strong molecular weight
barrier shields the CN-cycle processed regions. This occurs only after
the RGB bump in low-mass stars ($M \leq 2.3 M_\odot$) or on the early
AGB for the stars that do not go through the He flash ($M > 2.3
M_\odot$). Carbon depletion and nitrogen enhancement at the stellar
surface that would result from internal mixing are thus expected to
happen during these specific evolutionary stages according to the
proposed self-consistent scenarios \citep{charbon2000,charbon2010}. 
\subsection*{Paper outline}
In this paper, we propose a new re-analysis of the data available for
WGB stars and investigate the possible origin for their abundance
anomalies, giving particular attention to the lithium abundance
anomalies. In particular we investigate the possible physical
mechanisms that could account for the {\em simultaneous} lithium
overabundances and strong carbon underabundances. In the following, we describe the homogeneous redetermination of
atmospheric parameters and lithium abundances for the WGB stars known to
date in \S~\ref{sec:param} and \ref{sec:li}. We then place the WGB stars in the HR
diagram and make a tentative classification in terms of evolutionary
status and initial masses in \S~\ref{sec:evol}. In
\S~\ref{sec:abund} and \ref{sec:discussion} we finally explore possible scenarios to
consistently account for the light elements (Li, C, N and O)
abundances found in WGB stars based on the results of dedicated stellar evolution
models.

\section{Atmospheric parameters}\label{sec:param}
\begin{table*}
\caption{Basic data and atmospheric parameters of weak G-band stars, derived from Hipparcos catalog 
for HIP stars (col.\,2). Last column (H or A) means that the  atmospheric parameters have been derived
respectively thanks to the Houdashelt tabular data or the photometric calibration of Alonso (see text).  } 
\begin{tabular}{rccccccccccccccl}
\hline
   HD No & HIP No & m$_{v}$ &  B$-$V & $\pi$ &  $\sigma$(${\pi}$) & $\sigma$($\%$) &  log(L/L$_{\odot}$) & $\sigma$ & T$_{eff}$ & $\sigma$ &
log$g$ & $\sigma$ & Ref & \\
\hline
  18474   &     13965   &   5.47  &   0.869  &   6.390  &  0.350  & 5.4   &  2.192   &   0.088 &  5140
&  110  &  2.59  & 0.20  &   H & \\ 
  18636   &     13865   &   7.63  &   0.897  &   3.480  &  0.570  & 16.3  &  1.864   &   0.184 &  5080
& 110   & 2.90   & 0.29  & H & \\ 
  21018   &     15807   &   6.37  &   0.851  &   2.680  &  0.740  & 27.6  &  2.628   &    0.28 &  5300
&  80   & 2.21   & 0.39  & H & \\ 
  26575   &     19509   &   6.43  &   1.070  &   3.800  &  0.390  & 10.2  &  2.332   &   0.128 &  4690
& 110   & 2.29   & 0.24  & H & \\ 
  28932   &     21250   &   7.94  &   0.996  &   1.940  &  0.730  & 37.6  &  2.276   &   0.368 &  4880
&  90   & 2.42   & 0.48  & H & \\ 
  31274   &     22620   &   7.13  &   0.958  &   3.730  &  0.380  & 10.1  &  2.072   &   0.128 &  5040
&  80   & 2.68   & 0.24  & H & \\ 
31869   &-  &9.27   & 0.92  & -  & - & - & - & - & 
-&  - &  - &  -& -& \\
36552 & - & 8.08 & 0.84 &- &-  &-  &-  &-  &5250
& -  &1.5   & - & -& \\
  40402   &     28184   &   8.55  &   0.930  &   1.870  &  1.390  & 74.3  &  2.044   &   0.688 &  5010
& 120   & 2.70   & 0.80  & H & \\ 
  49960   &     32734   &   8.34  &   1.020  &   1.030  &  0.760  & 73.7  &  2.676   &    0.68 &  4830
&  80   & 2.00   & 0.79  & H & \\ 
  56438   &     35016   &   8.10  &   1.033  &   1.820  &  0.640  & 35.1  &  2.284   &   0.344 &  4790
& 110   & 2.38   & 0.45  & H & \\ 
  67728   &     39867   &   7.54  &   1.106  &   2.160  &  0.670  & 31.0  &  2.384   &   0.308 &  4660
&  90   & 2.23   & 0.42  & H & \\ 
  78146   &     44636   &   8.57  &   1.160  &   0.840  &  0.990  & 117.8 &  2.824   &   1.064 &  4530
&  90   & 1.74   & 1.17  & H & \\ 
82595 & - & 8.20 & 1.00 & -&  -&  -&  -&  -& 4910
&  - & 1.5  & - &- & \\
  91805   &     51816   &   6.09  &   0.940  &   2.870  &  0.430  & 14.9  &  2.736   &   0.172 &  5350
&  80   & 2.11   & 0.28  & H & \\ 
  94956   &     53561   &   8.45  &   0.960  &   2.580  &  0.790  & 30.6  &  1.816   &   0.308 &  4950
& 110   & 2.90   & 0.41  & H & \\ 
 102851   &     57724   &   8.78  &   1.053  &   1.400  &  1.220  & 87.1  &  2.244   &   0.796 &  4760
& 110   & 2.40   & 0.91  & H  & \\
 119256   &     66975   &   7.29  &   1.123  &   4.160  &  1.110  & 26.6  &  2.000   &   0.272 &  4870
& 400   & 2.69   & 0.38  & H  & \\
 120213   &     68009   &   5.95  &   1.411  &   4.070  &  0.330  & 8.1   &  2.644   &   0.112 &  4120
& 100   & 1.76   & 0.22  & A  & \\
120170 & - & 9.03 & 0.93 &  -&  -&  -&  -&  -& 5265
&  - & 2.70  & - & -& \\
124721 & - & 9.48 & 0.93 & - & - &-  &-  & - &
-&   -&   -&  -& -& \\
 146116   &     79604   &   7.68  &   1.070  &   1.160  &  0.710  & 61.2  &  2.856   &   0.572 &  4720
&  80   & 1.78   & 0.68  & H  & \\
 165462   &     88671   &   6.32  &   1.014  &   4.650  &  0.480  & 10.3  &  2.172   &   0.128 &  4840
& 110   & 2.50   & 0.24  & H  & \\
 165634   &     88839   &   4.55  &   0.938  &   9.620  &  0.260  & 2.7   &  2.276   &   0.064 &  5060
& 260   & 2.48   & 0.17  & H  & \\
 166208   &     88788   &   5.00  &   0.913  &   7.960  &  0.620  & 7.7   &  2.200   &   0.108 &  5050
& 110   & 2.55   & 0.22  & H  & \\
 188328   &     97922   &   7.02  &   0.584  &  12.110  &  1.670  & 13.7  &  0.932   &    0.16 &  6040
&  90   & 4.13   & 0.27  & H  & \\
 204046   &    105901   &   9.00  &   1.037  &   1.390  &  1.270  & 91.3  &  2.160   &   0.836 &  4790
& 110   & 2.50   & 0.94  & H  & \\
 207774   &    107885   &   8.92  &   0.969  &   2.320  &  1.300  & 56.0  &  1.720   &   0.528 &  4940
&  80   & 2.99   & 0.64  & H  & \\
 \hline
\end{tabular} 
\\
\footnotetext{}{H = Houdashelt 2000 }\\   
\footnotetext{}{A = Alonso et al 1999}       
\label{tab1}                                                                                              
\end{table*}                                                
 
The most complete list of weak G-band stars known up to now is
 presented in Table~\ref{tab1}. So far 28  WGB stars have been identified in the literature. This
represent a very small fraction ($< 1$ \%) of K giants and makes these objects
pertain to a very rare class. In order to study their position in the HR~diagram we have firstly determined their effective
 temperature \teff by using both $(B-V)$ and $(V-I)$ colors from the
 Hipparcos catalogue \citep{vanleeuwen2007} and the improved
 color-temperature relations for giants from \citet{houda2000} or
 \citet{alonso1999}.  Colors of some stars are affected by a little
 interstellar reddening and we take it into account for the
 determination of \teff and the bolometric correction $BC$ by
 using a least squares approximation in varying accordingly the color
 excess $E(B-V)$ and $E(V-I)$.  The error on \teff is
 estimated from our interpolation processing in the \citet{houda2000}
 tabular data .  The bolometric magnitude $M_\mathrm{bol}$ and gravity
 $\log(g)$ are given by :\\

\begin{equation}
M_\mathrm{bol} = V - 3.1\,E(B-V) + BC(V) + 5\log(\pi) + 5
\label{mbol}
\end{equation}

\begin{equation}
\log(g) = \log(M) +0.4M_\mathrm{bol} + 4\log(T_\mathrm{eff})
-12.505
\label{logg}
\end{equation}

where $\pi$ is the Hipparcos parallax \citep{vanleeuwen2007} and $M$ the mass of the star
in solar units. The $V$ magnitude is also taken from the release of
the Hipparcos catalogue by \citet{vanleeuwen2007}. In the preceding equation, gravity $\log(g)$ is little dependent on the mass of the star
in the range 2-4 solar masses. Due to the location of the stars in the HR diagram (see Fig.~\ref{fig1}) 
we assume $M$ = 3.5 \msun for the calculation of $\log(g)$ in Table~\ref{tab1}. The error on $\log(g)$ 
comes essentially from the error on \teff. 
The errors on the luminosities are computed using the quoted uncertainties in the parallaxes and
magnitudes. They may be huge for certain WGB stars as they propagate
from ill-determined parallaxes. 
Atmospheric parameters given in Table~\ref{tab1} are in excellent agreement with those calculated
recently by other authors \citep{takeda2008,wu2011}.

\section{Lithium Abundance}\label{sec:li}
Lithium abundance for most of the weak G-band stars, rederived from the available strengths of Li resonance 
line at 6707.8 \AA, are given in Table~\ref{tab2}.
The equivalent widths of Li line and metallicities are sourced from the literature 
\citep{hartoog1978,lambert1984}.
Solar metallicity is considered for the giants whose metallicity information was not available in the
literature.
Local thermodynamic equilibrium  (LTE) stellar model atmospheres 
computed by \citet{kurucz1994}\footnote[1]{www.kurucz.harvard.edu} with convection option on, and the revised radiative 
transfer code MOOG originally written by \citet{sneden1973} were used for the analysis. 
The well tested line list \citep{reddy2002} in the region of Li resonance line 6707.8\AA\ was adopted. 
Hyperfine structure and the $gf$ values of $^{7}$Li components of Li line at 6707.8\AA\ 
were taken from \citet{hobbs1999}. NLTE corrections are applied using the recipe given by \citet{lind2009}.

The information on the lithium abundance, in the Table~\ref{tab2}, is
given for 19 stars.  Lithium studies have not been attempted until now
on those nine stars whose Li information is missing in the
Table~\ref{tab2}.  Among the 19 stars with available lithium, 11 giants
can be qualified as ``{\em Li-rich}''. This denomination refers to an
unexpected large Li abundance compared to the post dredge-up value of
about 1.4 dex expected for giant stars at solar metallicity from
standard stellar evolution theory \citep[see this work and
][]{brown1989}.{\bf This represents about 39\% of the WGB stars and
almost 58\% of the WGB for which lithium has been searched for.}  The
remaining eight stars are lithium poor.  WGB stars present a large range
of Li abundances from -0.2 to 3.0 dex. Contrary to K giants, Li
overabundances are much more common in this subclass of stars, amounting
to about 58\% for those stars where the lithium region around 6707.8 \AA
has been observed. No weak G-band star is found to have Li abundance
more than main sequence or ISM value (A(Li)$\sim$ 3.28).  This very
large percentage by itself adds to the peculiarity of WGB stars among
other types of lithium-rich stars, in particular Li-rich giants that are
very rare.

We will discuss further the possible origin to these unexpected large
 lithium abundances in \S~\ref{sec:discussion}.

\section{HR Diagram}\label{sec:evol}
\begin{figure*}
\begin{center}
\includegraphics[width=0.9\textwidth]{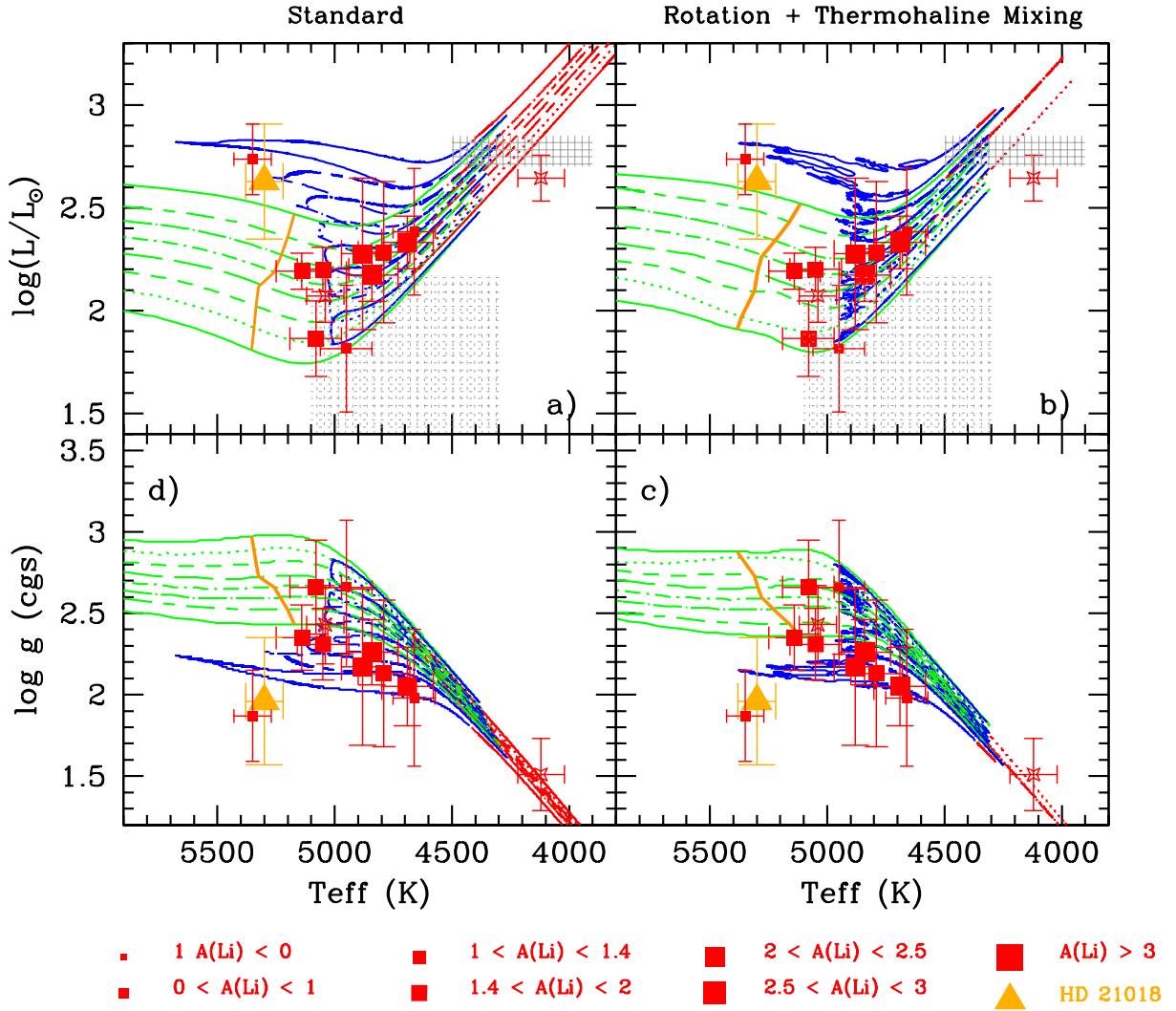}
\end{center}
\caption{Weak G-band stars with the best parallaxes (error on the log(L/L$_\odot$)
 lower than 0.5 dex) are plotted in the Hertzsprung-Russell diagram
 (upper panels) and in the (log g, T$_{eff}$) plane (lower panels) along
 with evolutionary tracks for models with initial masses  3, 3.2, 3.4,
 3.6, 3.8, 4.0, 4.2 and 4.5 M$_\odot$ as described in \S~4 (from bottom
 to top in the HR diagrams and in the reverse way, from top to bottom, in
 the (logg, \teff) diagrams.). Left panels
 represent classical models with no rotation nor diffusion processes
 included, and right panels show models including rotation and
 thermohaline mixing. The bold line intersecting the racks
 on the left side of the plots marks the beginning of the first
 dredge-up. The size of the symbols is proportional to the Li content
 (the more lithium, the larger the symbol). Empty symbols are those for
 which no lithium abundance exists to date. The triangle on each of the
 plots represents star HD 21018 which is a spectroscopic binary. The
 shaded areas in the HR diagrams indicate the location of the Li-rich
 K giants, corresponding to the bump area for models with initial masses
 lower than 2.3 M$_\odot$ and to the early-AGB phase for the models in
 the mas range presented here.}
\label{fig1}
\end{figure*}
In order to better evaluate the evolutionary status of the weak G-band
stars (also with respect to that of the Li-rich K giants), we have computed two grids of dedicated models
for stars with solar metallicity (the solar chemical composition adopted
as a reference is that of \citet{GN93}) in the mass
range 3 to 4.5 M$_\odot$. The models were all evolved from the pre main
sequence to the early AGB phase. They were computed using STAREVOL
V3.0 code, that has been extensively described in \citet{siess2006,palacios2006,decressin2009}. Standard models displayed in panels
a) and c) of Figure~\ref{fig1}, do not include mixing outside the
convective regions, and the
convective boundaries are fixed according to the Schwarzschild criterion
for the convective instability. No overshooting is included. The tracks associated with the rotating
models are displayed in panels b) and d) of the same figure. They were
computed assuming solid-body rotation in the convective envelopes (which
should result in a minimal estimate of the amount of shear mixing
possibly developing in the radiative region connecting the convective
envelope and the nuclear active regions ; see Palacios et al. 2006), and
rotation and thermohaline mixing were included from the Zero Age Main
Sequence up to the early-AGB phase. The formalism used is the same as in
\citet{charbon2010} and the reader is referred to this paper for
more details. Let us just recall that we assume an efficient
thermohaline mixing following \citet{ulrich1972}, which might turn out
to be a maximum efficiency approach given the lack of actual good description of this
thermodynamic instability in astrophysical regime \citep{cantiello2010}. As in \citet{charbon2010}, the rotation
velocity of the models when they reach the main sequence is assumed to
correspond to 45\% of their critical velocity, which results in mean
equatorial rotation velocities of about 200 km.s$^{-1}$. Mass loss has
been included in all models using the classical Reimers law \citep{reimers1975} up to the early AGB phase.

\begin{figure}
\begin{center}
\includegraphics[width=0.45\textwidth]{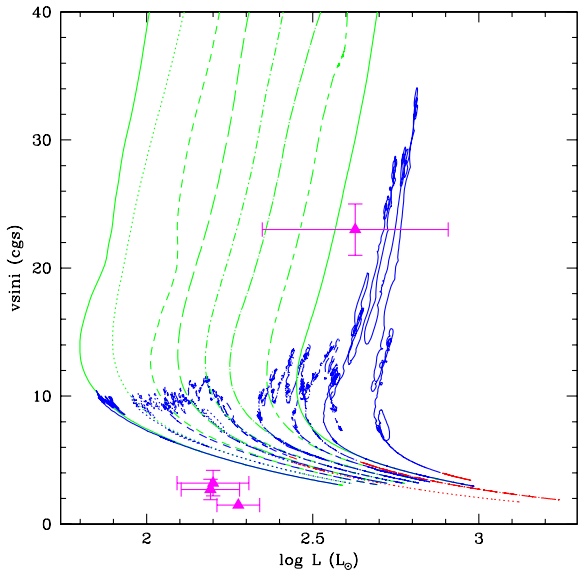}
\end{center}
\caption{ $\upsilon \sin i$ as a function of luminosity for the rotating
 models with initial masses from 3.0 to 4.5 M$_\odot$ (from left to
 right). The triangles are the observational values for HD 18474, HD
 21018, HD 166208, HD 188328 and HD 165634 as listed in
 Table~\ref{tab2}.}
\label{fig3}
\end{figure}

Figure~\ref{fig1} displays the luminosity and effective gravity of the
WGB stars as a function of the effective temperature together with the
evolutionary tracks. Only those objects with accurate-enough parallaxes
(e.g. $\epsilon$(log(L/L$_\odot$)) $<$ 0.5) are shown. The shaded areas
in panels a) and b) indicate the location in the HR diagram of the
Li-rich K giants as reported in \citet{kumar2011} and
\citet{charbon2000}. Only a small overlap exists and Li-rich WGB stars
being more luminous appear to clearly detach from the bulk of Li-rich K
giants, The thick line intersecting the tracks on the left part of the
plots indicates the beginning of the first dredge-up. {\bf WGB stars
appear clearly to be stars in the mass range 3 to 4.5 M$_\odot$ at the
clump. Their position in the HR diagram makes the determination of their
evolutionary status ambiguous : they could either be on the red giant
branch, undergoing the first DUP, or be in the core helium burning
phase. } \citet{cottrell1978} already suggested that WGB stars could be
core helium burning stars in their study of HD 91805, but contrary to
what they thought, WGB stars are not passed the helium flash since this
event only occurs in stars with a mass lower than approximately 2.3
M$_\odot$ \citep[see e.g.][]{palmerini2011}.  In the work Cottrell \&
Norris also suggested that meridional circulation induced by rotation in
main sequence stars of intermediate mass could be a possible mechanism
to explain the carbon strong depletion of WGB stars. We therefore
computed models including this transport process along with turbulent
shear and thermohaline mixing in order to test this hypothesis. \\
Rotation velocities ($\upsilon \sin i$) have been measured for a limited
number of WGB stars and are reported in
Table~\ref{tab2}. Figure~\ref{fig3} shows the evolution of the surface
equatorial velocity of our rotating models during the RGB and subsequent
evolutionary phases. In order to mimic a mean $\upsilon \sin i$
associated to what can be considered a mean equatorial velocity given by
1-D stellar evolution models, we have multiplied the velocity given by
our models by $\frac{\pi}{4}$, following \citet{chandrasekhar50}. The
observational data seem to suggest rotation rates lower (by a factor of
two approximately) than what is predicted by our models. Let us mention
here that conservatively adopting a solid-body rotation law for the
convective envelope favours larger surface rotation rates. Using a
probably more realistic differential rotation in the extended convective
envelopes of these stars as they evolve through the giant phases will
lead to lower surface rotation velocities \citep[see][]{brun2009}. HD
21018 is on the other hand quite well reproduced considering the large
error bars on the parallax. The very slow velocities estimated for the
five stars quoted in Table~\ref{tab2} are consistent with what could be
expected from stars undergoing core helium burning.

As can be seen from Fig.~\ref{fig1}, the evolutionary tracks are slightly modified by
the action of rotation and thermohaline mixing, having less extended and
noisier blue loops. The tracks nonetheless match the region occupied by
the WGB stars as well as the standard models do.\\

In the following, we confront the models, standard and with rotation, to
the abundance data and discuss the possible origin of the WGB stars in
the light of this confrontation. 

\section{Abundances}\label{sec:abund}
\begin{table*}
\caption{Chemical abundances and rotational velocities of weak G-band stars}
\begin{center}
\begin{tabular}{rcccccccccc}
\hline
  HD   &    [Fe/H] & A(Li) & W$_{\lambda}$ & A(Li) &
 A(Li)$_{NLTE}$ & [C/H]$^{b}$ & [N/H]$^{b}$ & [O/H] & \ciso & \vsini\\
\hline
 18474 &   $-$0.10$^{a}$ &  1.42$^{a}$  & 29$^{a}$ & 1.33 & 1.48 &-1.6 & 0.2 &-&- & 1.9$^{g}$\\
 18636 &   $-$0.11$^{b}$ &1.77$^{b}$ & 66$^{b}$ & 1.71 & 1.853 &-&-&-&- &- \\
 21018$^{p}$ &    - & 2.89$^{c}$ & 245$^{c}$ & 3.13  & 3.06&-&-&-&-&20.1$^{g}$ \\
 26575 &     $-$0.01$^{b}$ & 2.02$^{b}$ & 239$^{b}$ & 2.16 & 2.34&-&-&-&-&- \\
 28932 &     - & 2.25$^{a}$ & 158$^{a}$ & 2.04 & 2.212 &-&-&-&-&-\\
 31274 &  $-$0.27$^{b}$ & $< 0.52^{b}$& $< 4^{b}$   & $-$0.67 &-&-&-&-&-  \\
 31869 & -&- &- &-  &- &-&-&-&- &- \\
 36552 &  $-$0.17$^{b}$  &   2.58$^{b}$ & 173$^{b}$ & 2.64 & 2.70 &-&-&-&-&-\\
 40402 &  $-$0.45$^{b}$ &  2.44$^{b}$ & 215$^{b}$ & 2.60 & 2.586 &-&-&-&-&-\\
 49960 &  -&- &- &-  &- &-&-&-&- &- \\ 
 56438 &  $-$0.38$^{b}$ & 1.43$^{b}$ & 104$^{b}$ & 1.59 & 1.774 &-&-&-&-&-\\
 67728 &-   & 0.76$^{a}$ & 15$^{a}$ & $-$0.22 & 0.06 &-&-&-&-&-\\
 78146 &  $-$0.31$^{b}$ & 1.01$^{b}$ & 22$^{b}$ & $-$0.27 & 0.01 &-&-&-&-&-\\
 82595 &  $-$0.10$^{b}$ & 1.14$^{b}$ & 22$^{b}$ & 0.89 & 1.136 &-&-&-&-&-\\
 91805 &  0.0$^{e}$ & 0.25$^{b}$ & 03$^{b}$ & 0.06 & 0.202 & -1.4 & 0.6 & 0.1$^{e}$ &-&-\\
 94956 &  - & 1.07$^{a}$ & 15$^{a}$ & 0.65 & 0.834 &-&-&-&-&-\\
102851 &  -&- &- &-  &- &-&-&-&-&-  \\
119256 &  -&- &- &-  &- &-&-&-&- &- \\
120213 &  -&- &- &-  &- &-&-&-&- &- \\
120170 &  $-$0.07$^{a}$ &  3.03$^{a}$ & 253$^{a}$ & 3.14 & 3.02 &-&-&-&-&-\\
124721 & -&- &- &-  &- &-&-&-&- &- \\
146116 & -&- &- &-  &- &-&-&-&- &- \\
165462 & -   & 2.05$^{a}$ & 141$^{a}$ & 1.87 & 2.068 &-&-&-&-&-\\
165634$^{p}$ &  $-$0.15$^{f}$ &  0.75$^{f}$ &- &- &- &-1.49 &0.23 &-0.38$^{e}$&4.1$^{e}$ & 1.49$^{h}$\\
166208$^{p}$ &  +0.2$^{a}$ & 1.65$^{a}$ & 74$^{a}$ & 1.73 & 1.90 & -0.9 & 0.8 &- &4.5$^{d}$&2.9$^{g}$\\
188328 &  -&- &- &-  &- &-&-&-&-&6.3$^{g}$  \\
204046 &  -&- &- &-  &- &-&-&-&-&-  \\
207774 &  - & 1.5$^{a}$ &  30$^{a}$ & 1.06 & 1.24&-&-&-&-&- \\
\hline
\end{tabular}\\
\end{center}
\footnotetext{a}{$^{a}$ \citet{hartoog1978} },
\footnotetext{b}{$^{b}$ \citet{lambert1984} },
\footnotetext{c}{$^{c}$From our spectra},
\footnotetext{d}{$^{d}$ \citet{day1980} },
\footnotetext{e}{$^{e}$ \citet{cottrell1978} },
\footnotetext{f}{$^{f}$ \citet{sneden1978} },
\footnotetext{g}{$^{g}$ \citet{glebocki2005} },
\footnotetext{h}{$^{h}$ \citet{hekker2007} },
\footnotetext{p}{$^{p}$Binary},
\label{tab2}
\end{table*}

\begin{figure*}
\begin{center}
\includegraphics[width=0.9\textwidth]{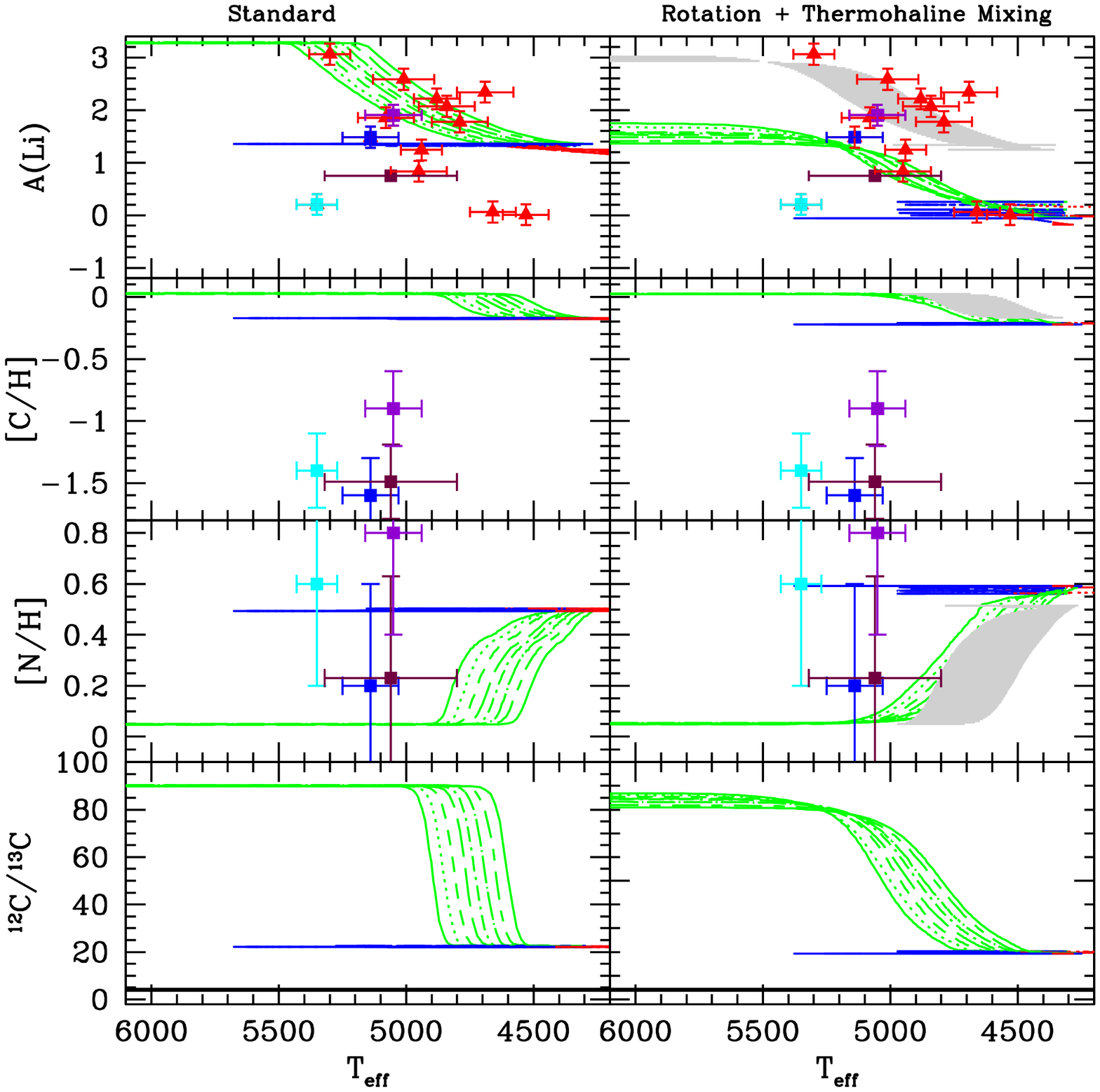}
\end{center}
\caption{Abundances of lithium, carbon, nitrogen and carbon isotopic
 ratio as a function of the effective temperature for the objects listed in
 Table~\ref{tab2}. A conservative error bar of 0.2 dex has been adopted for the
 NLTE lithium shown in the upper panels. The tracks are for
 standard (left column) and rotating models (right column) with masses
 ranging from 3.0 to 4.5 M$_\odot$ (from left to right). The colors of
 the tracks indicate the evolutionary phase : black for main sequence,
 green for RGB, blue for central He burning and red for AGB
 phase. Squares in different shades of blue are used in the upper panel
 to represent those stars for which [C/H] and [N/H] is also available
 from the literature (see text). On the lower panels, the thick black
 line at $^{12}$C/$^{13}$C = 4 represents the carbon isotopic ratio found in
 HD 16634 \citep{sneden1978} and in HD 166208 \citep{day1980}. The grey
 areas in the left column plots show the region occupied by tracks for
 rotating models with $\upsilon_{\rm ZAMS} = 50$ \kms (see text).}
\label{fig2}
\end{figure*}

The position of WGB stars in Fig~\ref{fig1} indicates that they are
bright stars of intermediate mass. The expected evolution of the
surface abundances of lithium, carbon, nitrogen and carbon isotopic
ratio for our standard and rotating models is shown in
Fig.~\ref{fig2} together with the values reported in Table~\ref{tab2}
for the WGB stars. In this figure, the squares in different shades of
blue/violet represent correspond to HD 166208 (violet), HD 165634 (brown), HD
91805 (cyan) and HD 18474 (blue).

\subsubsection*{Lithium}
The evolution of the surface abundance of lithium in standard (left) and
rotating models (right) is displayed in the first row. The weak G-band
stars are either Li-rich or Li depleted.\\
For those stars that have Li abundances lower than 1.4 dex, we clearly see that standard stellar evolution models fail to
reproduce them, the level reached after the first and second dredge-up
episodes being of about 1.4 dex. On the contrary {\bf the models including
rotation-induced mixing (e.g. meridional circulation + secular turbulent
shear instability) and thermohaline mixing undergo lithium depletion
already during the main sequence phase and achieve A(Li) $\approx 0$ in
very good agreement with the NLTE corrected lithium abundance derived
for HD 67728, HD 78146 and HD 91805.}\\

The rest of the WGB stars exhibit lithium abundances that would be typical of
first dredge-up dilution in standard (non-rotating) models and in models
with slow rotation on the ZAMS (50 \kms). This can be seen from the upper
panels of Fig.~\ref{fig2}, where the grey shaded area corresponds to the
location of tracks for slow rotating models. Among those for which
lithium lines were actually observed, the proportion of WGB stars with
$A(Li) > 1.4$ dex is of about 58\%, which means either that a large
proportion of core helium burning WGB stars are Li-rich, or that a large
fraction of the WGB stars are young red giants in which the first
dredge-up is not completed yet.\\ We postpone a more detailed discussion
to \S~\ref{sec:discussion}.

\subsubsection*{Beryllium}
The evolution of the surface Be abundance in our models lead to a mean
post-dredge-up value (first and second) of about -0.2 dex for the models
computed, while the models including rotation and thermohaline mixing
reach A(Be) $\approx -0.65$ dex. These values are to be compared with
the abundances estimated from IUE spectra by \citet{partha1984} of about
-0.5 dex. Errorbars on observed values are not detailed in the original
paper but the value is considered an upper limit so that non-standard
models are here again favoured.

\subsubsection*{Carbon}

Actual derivation of carbon abundances exist only for a bunch of stars
with known parallaxes, that are represented together with our models
predictions in the second row of Fig.~\ref{fig2}. While models including
rotation-induced and thermohaline mixing lead to [C/H] $\approx -0.25$
dex by the end of the core helium burning phase, about 0.05 dex larger a
depletion than predicted for classical stellar evolution models, these
values are far from reaching the extreme underabundances derived for HD
166208, HD 165634, HD 91805 and HD 18474 (see also
Table.~\ref{tab2}). \\ The combined action of meridional circulation and
turbulent shear and thermohaline mixing that has been proposed as a
physical process to explain light-elements (Li, C and N) abundance
anomalies in RGB stars \citep{charbon2010} clearly fails here to account
for very large carbon depletion in WGB stars when using the same set of
parameters as those adopted for lower mass stars. Let us note that among
the WGB stars for which [C/H] is available, HD 91805 is ``normal'' in
terms of lithium (and also nitrogen, see below) abundance when compared
to non-standard models including rotation and thermohaline mixing, while
the other three stars are Li-rich ($A(Li) \geq 1.4$ dex). \\
{\bf The carbon deficiency, the common feature to the WGB stars, thus seems to be decorrelated from the other
abundance anomalies, in particular the lithium content, and could
have a different origin.}

\subsubsection*{Nitrogen}

Nitrogen abundances are available from the literature for the same stars
for which the carbon abundance has been derived. The third row (from the
top) of Fig.~\ref{fig2} displays the evolution of the [N/H] as
predicted by stellar evolution and the abundances derived for HD 166208,
HD 165634, HD 91805 and HD 18474.  
In our non-standard models, nitrogen is
enhanced during the RGB phase (up going parts of the tracks) as a result
of the continuously ongoing rotation-induced mixing allowing to
bridge the gap between the convective envelope and the hydrogen burning
shell. Thermohaline mixing only develops near the core, but anyway
remains much slower than meridional circulation and shear-induced
turbulence, and is not the lead process shaping the surface abundance
pattern of our models.\\

Both standard and non-standard models
lead to surface abundances at the core helium burning phase that
are in agreement with the observational data within the (large) error
bars. We may however distinguish two groups :
\begin{enumerate}
\item the more nitrogen rich stars HD 91805 (cyan square at $T_{eff} = 5350 K$)  and HD 166208
(purple square at $T_{eff} = 5050 K$) are clearly better fitted by
{\bf \em rotating models undergoing core helium burning}, in particular HD 91805;
\item  the nitrogen abundances of HD 165634 and
HD 18474 fall in between [N/H]$_SGB$ ($\approx 0.05$ dex) and
[N/H]$_{\rm clump}$ ($\approx 0.6$ dex) as predicted by both our sets of
models. This is quite puzzling and definitely not of any help to pinpoint the
evolutionary status of these stars. 
\end{enumerate}

\section{Origin of the peculiar chemical pattern of WGB stars}\label{sec:discussion}

From the preceding sections \S~\ref{sec:evol} and \S~\ref{sec:abund},
weak G band stars appear to be intermediate mass stars with initial masses in the
range 3 to  4.5 \msun that are undergoing either the first dredge-up
or are in the core helium burning stage. A large
fraction of these stars is lithium rich (A(Li) $>$ 1.4 dex), and all of
them present large carbon underabundances even when compared to non-standard stellar evolution models including
thermohaline mixing and rotation. \\
When trying to understand the observations, one needs to be able to
account for the following features :
\begin{enumerate}
\item All WGB stars are carbon deficient, and for the 4 stars with
      available measurements for the carbon abundance, [C/Fe] $< -1$
      dex; carbon isotopic ratio is also seemingly very small (around
      equilibrium value 4). The observed underabundance are larger by at
      least 1 dex than the one predicted by standard and non-standard models;
\item A large fraction (about 58\%) of the WGB stars for which lithium
      lines have been observed exhibit $A(Li) > 1.4$ dex;
\item Lithium and nitrogen surface abundances as well as evolutionary
      status are well matched by stellar evolution models including
      rotation and thermohaline mixing for those stars that present a low lithium surface abundance; 
\item The rotation velocity of WGB stars with measured  $\upsilon \sin i$ is
      marginally reproduced by rotating stellar evolution models with
      $\upsilon_{\rm ZAMS} \approx 200$ \kms.
\end{enumerate}

\subsection{''Li-poor'' weak G band stars}

As partly discussed in \S~\ref{sec:abund}, lithium poor WGB stars are
actually normal except for their carbon abundance. HD 91805 is used as a
prototype since we have access to its luminosity, temperature, gravity,
lithium, carbon and nitrogen abundances. This star is well fitted by
models including rotation, rotation-induced mixing (meridional
circulation + shear turbulence), and thermohaline mixing with an initial
mass of about 4.5 \msun. \\

If we consider an internal origin for the carbon underabundance, as
proposed by \citet{cottrell1978,tomkin1984}, the evolutionary status of
red giants (RGB) stars should be ruled out.
Considering their mass, WGB stars experience the end of the first
dredge-up while already undergoing core helium burning, and no such
extra mixing processes as rotation-mixing or thermohaline mixing are
expected (and actually seen in the models) to be
efficient enough to create the large carbon depletion. Furthermore, admitting that some extra-mixing acts as
an overshooting and makes the dredge-up deeper, the very large extent of
the convective envelope at this stage implies a very large dilution
factor that will prevent the carbon abundance and carbon isotopic ratio
to drop to the observed value within the duration of the RGB
phase. Even though the mixing reaches down to the region where CN-cycle
occurs, this region is far too shallow compared to the convective
envelope for this late region to be completely processed. \\
Meridional circulation and shear turbulence are actually efficient in
earlier evolutionary phases as can be seen in Fig.~\ref{fig2}, but they
mainly act on the lithium abundance and the associated mixing does not
reach the carbon depleted regions that are located deeper in the
core. Contrary to what was suggested by \citet{lambert1984} or
\citet{cottrell1978}, its seems very unlikely that carbon depletion is
the result of internal mixing during the early evolution of the
progenitors of WGB stars. \\

During core helium burning, on the other hand, the HBS crosses the
composition gradient left by the dredge-up when the stars are located at
the ``turning point'' of the blue loop, and then the regions were
CN-cycling occurs are again accessible to extra-mixing processes. However, the
physical distance separating the nucleosynthetic regions from the now
much shallower convective envelope is very large (more than 22 \rsun) : for an
extra-mixing process to be efficient it is then needed that its
characteristic timescale would be shorter than the evolutionary
timescale of about 20 to 200 Myrs typical for stars in the mass range
considered here. In terms of diffusion coefficient, this would imply
$D_{\rm extra mix} \approx 10^8 - 10^{10}$ cm$^2$.s$^{-1}$. The total
diffusion coefficient obtained in our non-standard models ($D_{\rm MC} +
D_{\rm turb} + D_{\rm Thl}$) is always smaller by at least one order of
magnitude. Such high rates of mixing are difficult to obtain using the
available prescriptions for hydrodynamical instabilities.\\ 

Altogether these points indicate that the extra-mixing processes
included so far in our stellar evolution models (meridional circulation
+ turbulent shear + thermohaline mixing) cannot produce the huge carbon
underabundances but actually well reproduce the lithium and nitrogen abundances
measured for 3 of the WGB stars if likely rotation velocities are
adopted for their main sequence progenitors.

\subsection{Li-rich weak G band stars}

As already mentioned several times, WGB stars with $A(Li) \geq 1.4$
dex represent a large percentage of these stars. Such lithium
abundances will be considered an evidence for lithium overabundances if
the stars are evolved past the first dredge-up, but would be within the
expected range for stars that are currently undergoing the first
dredge-up.\\ Let us analyze the available observational data using both
interpretations and the consequences of adopting one or the other point
of view for the origin of the peculiar abundance patterns of WGB stars.

\subsubsection*{WGB stars as core helium burning stars}

If one assumes from their location at the
clump in the HR diagram and the discussion in the previous
paragraph concerning the evolutionary status of the Li-poor ones, that
WGB stars are core helium burning stars, then lithium enrichment is everything but an
exceptional feature  among this stellar subclass.  The Li-rich stars are rare among other classes of
giants: in the case of low-mass stars, they are the result of seemingly very short enrichment
episodes at which we manage to catch a glimpse for only 1\% - 5\% of the
RGB stars \citep{charbon2000,palacios2001}. For more massive stars, they are produced by Li-enrichment
episodes in stars with M$_\star \geq 5 \msun$ that undergo Hot Bottom
Burning during the thermal pulse AGB phase \citep{palmerini2011}. \\
Weak G band stars do not appear to be on the TP-AGB phase (too warm and
not luminous enough).  During the core helium burning phase there is no such thing as Hot
Bottom Burning, and to explain Li-enrichment by internal mixing, one
should invoke Cool Bottom Processing, a general denomination for any
unspecified mechanism able to connect the convective envelope with
deeper regions where nuclear reactions occur.\\

Cool Bottom Processing in relation with rotation-induced hydrodynamical
instabilities has been proposed to account for Li enrichment at the RGB
bump \citep{palacios2001} in the lithium flash scenario. In this
scenario, the convective envelope of the RGB star is enriched with
lithium when the outer regions of the HBS were $^7{\rm
Li}(p,\alpha)^4{\rm He}$ occurs, becomes convectively unstable and
connects with the envelope. This episode is very prompt in time, and is
rapidly followed by a deepening of the extra-mixing. Carbon and lithium
abundances subsequently decrease at the surface. \\ While one could be
tempted to transpose this scenario to the Li-rich WGB stars, a
difficulty arises since it excludes a {\em simultaneous} increase of Li
{\em and} depletion of carbon. This could be overcome if we consider that
the stars already present a strong carbon depletion prior to the red giant phase. Another
difficulty resides in the timescales : short-time enrichment is
difficult to reconcile with when almost 40\% of the stars in the WGB
class being Li-rich, but keeping in mind that WGB stars are very
rare {\em per se}, this difficulty might not be as difficult to overcome
as it may appear at first sight. \\ 
\citet{lambert1984} suggested that the abundance anomalies of WGB stars,
including Li enrichment and strong carbon depletion could be due to
diffusion in the early stages of the evolution of A stars. They
speculate that WGB stars could be the progeny of Ap stars which are
well known magnetic CP stars. The role they attribute to magnetism is
not at all developed, as they only use the fact that Ap stars are rare
and that the paucity of WGB stars is consistent with such a
filiation. Considering the masses derived for the WGB stars in
\S~\ref{sec:evol}, they cannot be descendants of A stars and are the
progeny of late-B type stars. If we still consider the possibility that
WGB stars are descendant of magnetic Bp stars, we can expect in these
stars effects similar to those in magnetic Ap stars and a similar
evolution. Concerning a diffusion scenario including the combination of
gravitational settling and radiative accelerations, any abundance
stratification that would be built during the main sequence will be
erased by the first dredge-up (O. Richard, private
communication). Moreover, for stars with \teff $> 12000$ K on the main
sequence, the effects of diffusion are expected to be essentially
confined to the atmosphere, and non-magnetic evolution models by \citet{turcotte2003}
show that no specific accumulation or depletion of carbon is
expected in the interior of these stars that could appear at their
surface as they will evolve up the giant branch.\\
The evolutionary models for
Ap/Bp stars are not yet available since the effects of magnetic fields and
how to take them into account is still a matter of debate. Still,
\citet{charbon2007b} suggested that magnetic Ap stars give rise to the
very few RGB stars that present light elements abundances ($^3{\rm He},
^7{\rm Li}$, C, N) complying to the predictions of standard and rotating
models. Strong magnetic fields are actually invoked as a mean to inhibit the
thermohaline mixing that is thought to be responsible for the light
elements abundance variations seen at the surface of RGB stars that have
evolved past the bump. Following this work and considering that for
intermediate mass stars, thermohaline mixing is inefficient unless maybe
during the TP-AGB phase, that is at a more advanced evolutionary stage
than that attributed to WGB stars, magnetic fields are not expected at all to
favour carbon depletion of any sort, and the scenario proposed by \citet{lambert1984} is
discarded.\\

Mass transfer (or mass accretion) appears as one of the very few options left to try and
understand the WGB phenomenon. 
From the HST UV  spectra \citet{bohm2000}
find evidence for the presence of a white dwarf companion to the weak G-band
star HD165634.  On the other hand \citet{tomkin1984} searched for
binarity in a sample of 7 WGB stars of the northern hemisphere, finding a
binarity rate of 15\%-40\% , that is not very different from that
expected for normal K giants. This result is however questionable
since the number of targets selected was rather small and these observers did not exhibit the persistence that is really
needed for an investigation of radial velocity variability because they stopped
their observations after only two years \citep[see arguments also given by][]{griffin1992}.\\
For the binary scenario to work, the secondary of the system should also
be a star from which the accreted matter would be carbon depleted (and
possibly sometimes {\em also} lithium enriched). According to
\citet{forestini1997} (see their Table 8), stars in the mass range 5-6
\msun at solar metallicity are expected to produce yields baring such a
chemical imprint. In their models, due to the operation of efficient
Hot Bottom Burning, they predict that the yields of
$^7{\rm Li}$ are positive and more interestingly, that the net yields of
$^{12}{\rm C}$ are negative with values up to -1.21 $10^{-2}$
\msun. \citet{forestini1997} do not give any expected decrease of C
at the surface of the secondary. They find lifetimes of 111 Myrs and
65.8 Myrs for their 5 \msun and 6 \msun respectively at solar
metallicity, to be compared with the 473 Myrs and 210 Myrs for their
3\msun and 4 \msun respectively. These numbers imply that the mass
accretion from the more massive shorter lived primary onto the secondary
would occur during the early main sequence evolution of the latter. As
main sequence stars in the mass range 3 to 4.5 \msun have a radiative
envelope, it is likely that the material that would be accreted would remain at the
surface and possibly slowly diffuse inwards. If WGB stars are intermediate-mass
stars undergoing core helium burning, they have already evolved through
the first dredge-up, one of the main signatures of which is the decrease
of surface abundances of carbon and lithium, and the increase of
nitrogen abundance. The dredge-up of carbon depleted, lithium enhanced
material could exacerbate the carbon depletion at the end of the
dredge-up (but this need to be actually computed and check before
drawing any conclusion) but will certainly not affect the lithium
abundance decrease since the point is that the base of the convective
envelope enters regions where lithium is destroyed by proton
captures. \\
Another uncertainty of the binary scenario is that of the
yields of intermediate-mass stars. They indeed vary greatly from one source to another \citep[see for
instance][]{karakas2007}, making any conclusion illusive.

\subsubsection*{WGB stars as young red giant stars (SGB/RGB) }

Contrary to the Li-poor WGB stars, that are very seemingly rotating core He
burning stars with anomalous carbon depletion, it is much more difficult to actually assess the
evolutionary status of those that exhibit a lithium abundance larger
than that expected after the completion of the first dredge-up. \\
In fact, if no postulate is made concerning a common evolutionary status
for all the stars populating the weak G-band subclass, then a strong
ambiguity arises. From the sole position in the HR diagram and the
comparison with the stellar evolution predictions for lithium abundance
(and marginally nitrogen), we can not rule out the possibility that the
so-called ``{\em Li-rich}'' WGB stars could turn out to be just ``{\em normal}'' stars as far as lithium is concerned, meaning with lithium abundances
consistent with models predictions, according to their evolutionary
status and their rotational history.\\ Indeed, from
Fig.~\ref{fig2} we see that the stars with large lithium abundance are well fitted by the standard tracks of
intermediate mass stars that are undergoing the first
dredge-up. According to these tracks, they should thus not be considered
as {\em lithium rich} since the first dredge-up episode is not
completed.\\

We may argue that the WGB stars should have experienced
some rotational mixing being the progeny of B-type stars.  The initial rotation spread observed in main sequence B-type stars in
the mass range 2 to 4 M$_\odot$ as shown by \cite{HGMcS2010} is actually
very large, and it might well be that WGB stars are also the descendants
of slow rotating late-B type stars. \\ In order to investigate the
effect of a slower rotation velocity of the progenitors of the WGB stars
on the ZAMS, we have computed 3 models (with initial masses of 3.0 \msun
3.5 \msun and 4.5 \msun) for which we adopted $\upsilon_{\rm ZAMS} = 50$ \kms. The domain covered by these slow rotators is represented by
the shaded area in Fig.~\ref{fig2}.\\ $\upsilon_{\rm ZAMS} = 50$ \kms
corresponds to a ratio $\upsilon_{\rm eq}/\upsilon_{\rm crit} \approx
0.1$. With such a slow rotation, the transport due to meridional
circulation and turbulent shear instability is not very efficient during
the main sequence evolution so that lithium is little depleted at the
surface of our models when they reach the turn-off. The deepening of the
convective envelope during the first dredge-up leads to a decrease of
the surface lithium and carbon abundances and to an increase
of the nitrogen abundance similar to that obtained for standard stellar
evolution. \\

Although it has been commonly assumed that WGB stars
shared the same evolutionary status, our models along with the chemical constraints (from lithium and nitrogen) that are available
for WGB stars could be telling a different story. This subclass
could be populated with
the progeny of slow to mild rotating ($\upsilon_{\rm eq}/\upsilon_{\rm
crit} \leq 0.45 $) B-type stars that are either experiencing
the first dredge-up or that are past this phase and are core helium
burning stars. \\ In this context the large carbon underabundance
associated to the G band weakness observed in all these stars, and the
narrow mass range become the only common features of WGB stars. \\ As
mentioned earlier a binary scenario is not to be excluded, but would
deserve an in depth study before any conclusions are drawn.\\

\section{Conclusions}

In this study we have made a tentative re-analysis of the data existing
for the WGB stars, in particular concerning fundamental parameters and
lithium abundances. We have confirmed that WGB stars are within the mass
range 3-4.5\msun. However, the too few 
abundance determinations of Li, N and C and the fact that the WGB stars are
located at the clump where stellar evolution tracks corresponding to
the subgiant, red giant and core helium burning phases are intertwined
prevent us to assess any sound evolutionary status to the
WGB stars. Concerning the chemical peculiarities, we have used dedicated
stellar evolution models to test their endogenous origin, and shown
that for those WGB stars that are not Li-rich, lithium and nitrogen
abundances can be reproduced by rotating models . On the other hand the carbon deficiency of all WGB
stars is difficult to reconcile with an internal origin and we propose that it is
decorrelated from the other chemical peculiarities. The case of
lithium is still under debate as it is not clear, because of the
degeneracy of the evolutionary status, whether the WGB stars with $A(Li)
\geq 1.4$ dex are actually Li-rich (core helium burning stars) or normal
(subgiant or RGB stars undergoing the first dredge-up). \\
Carbon depletion could be associated to an early mass transfer from a
more massive (5-6 \msun) companion under specific hypothesis, but yields
of intermediate mass stars in that mass range are very uncertain and no
firm conclusion should be drawn. As the rate of binarity among
WGB stars is far from clear, and is at least similar to that in normal
giants, a new search for duplicity would be helpful.\\
It is quite clear from this study that more data are needed to shed
light on the weak G-band stars puzzle.
Considering the work by \citet{miglio2011} on the use of asteroseismic data
to lift the status degeneracy between giants and clump stars, applying
such an analysis to the oscillation spectra of WGB stars could be a very
important step
to clarify the evolutionary status of WGB stars.\\ Actually deriving proper
abundances of light elements for these stars is also needed to
start and better understand these overlooked puzzling stars.

\section{Acknowledgements}

The authors thank the referee for very thoughtful comments that have
helped improve and clarify the work presented here. MP is thankful to Prof. Shoken Miyama, Prof. Ramanath Cowsik and Prof. Yoichi Takeda
for their kind support, encouragement and hospitality. AP thanks
Dr. O. Richard  and Dr. F. Martins for fruitful discussions
on diffusion and modelling.\\

\appendix
\section{Notes on Individual Objects}
\subsubsection{HD~21018}
It is a well-known visual binary with an orbital period of 287 days. It has highest Li among all the weak G-band stars. Charbonnel (2000) suggested that
it is undergoing the first dredge-up dilution.  
Due to the duplicity, the position of this object in the HR~diagram may be uncertain.

\subsubsection{HD~165634}
It is reported to have white dwarf binary companion from the IUE spectral studies by \citet{bohm2000}.
Li information is not available for this star.

\subsubsection{HD~188328}
It is a well-known visual binary ; both components are separated by about $2\arcsec$ according to
\citet{douglass2000}.
The position of this object in the HR~diagram occupies near the main-sequence turn-off.
Due to the visual binarity the position in the HR~diagram may be uncertain.
Li information is not available.

\subsubsection{HD~120213}
It may be a binary star. It is classified as a weak G band star in \citet{hartoog1977}. 
Further studies on this star classified as 
Ba star of class 0.5 \citep{lu1991}. In general,
most of the Ba stars are suggested to be binaries.

\subsubsection{HD~166208}
HD~166208 is a spectroscopic binary with a period of 5,5 years, low amplitude (3km.s$^{-1}$), and
moderate eccentricity (0.4) \citep{griffin1992}. \citet{griffin1992} emphasizes the point that the
eccentricity of the orbit of HD~166208 and its small mass function make distinctly less likely
that in the cases of barium stars that there has been transfer of material between the components
during a giant-phase evolution of the companion.

For few stars, like HD~36552, HD~82595, and HD~120170, luminosities are not derived due to the lack of 
Hipparcos parallaxes.
There is no basic data for HD~124721 and HD~31869 available in the literature.

\bibliography{Palacios2011_WGB}

\end{document}